\documentclass[journal]{vgtc}                     

\usepackage{amsmath,amssymb,amsfonts}
\DeclareMathOperator{\diag}{diag}
 \usepackage{siunitx} 

\onlineid{0}

\vgtccategory{Research}

\vgtcpapertype{theory/model}

\title{Modeling online adaptive navigation in virtual environments based on PID control}

\author{%
  \authororcid{Yuyang Wang}{0000-0003-0242-8935},  
   \authororcid{Jean-Rémy Chardonnet}{0000-0002-8926-1359}, and
  \authororcid{Frédéric Merienne}{0000-0003-4466-4776}
}

\authorfooter{
  \item
  	Yuyang Wang is with Hong Kong University of Science and Technology (Guangzhou).
        E-mail: yuyangwang@ust.hk
   \item Jean-Rémy Chardonnet and Frédéric Merienne are with Arts et Métiers Institute of Technology, France.
        E-mail: {jean-remy.chardonnet | frederic.merienne}@connect.hkust-gz.edu.cn
}

\abstract{%
It is well known that locomotion-dominated navigation tasks may highly provoke cybersickness effects. Past research has proposed numerous approaches to tackle this issue based on offline considerations. In this work, a novel approach to mitigate cybersickness is presented based on online adaptative navigation. Considering the Proportional-Integral-Derivative (PID) control method, we proposed a mathematical model for online adaptive navigation parameterized with several parameters, taking as input the users' electro-dermal activity (EDA), an efficient indicator to measure the cybersickness level, and providing as output adapted navigation accelerations. Therefore, minimizing the cybersickness level is regarded as an argument optimization problem: find the PID model parameters which can reduce the severity of cybersickness. User studies were organized to collect non-adapted navigation accelerations and the corresponding EDA signals. A deep neural network was then formulated to learn the correlation between EDA and navigation accelerations. The hyperparameters of the network were obtained through the Optuna open-source framework. To validate the performance of the optimized online adaptive navigation developed through the PID control, we performed an analysis in a simulated user study based on the pre-trained deep neural network. Results indicate a significant reduction of cybersickness in terms of EDA signal analysis and motion sickness dose value. This is a pioneering work which presented a systematic strategy for adaptive navigation settings from a theoretical point. 
}

\keywords{PID Control, VR, Navigation}

\graphicspath{{figs/}{figures/}{pictures/}{images/}{./}} 

\usepackage{tabu}                      
\usepackage{booktabs}                  
\usepackage{lipsum}                    
\usepackage{mwe}                       

\usepackage{mathptmx}                  

\begin{document}


\firstsection{Introduction}

\maketitle

Thanks to increasing computing power and the availability of many affordable head-mounted displays (HMDs) such as HTC Vive and Oculus Quest, the word ``metaverse'' has aroused profound discussion on the application of VR technologies among the public including mass media, industry, and academic community. Engineers are able to develop dozens of applications including training for medical operations, the organization of 3D virtual conferences, playing immersive games, visualizing 3D models, \textit{etc.}~\cite{lee2021all}. When exposed to immersive environments, users can easily perform many tasks in a virtual world after wearing VR glasses, including walking/running, shooting, and fighting. The accomplishment of these tasks requires users to navigate through multiple virtual environments, which could be the most fundamental interaction process~\cite{Wang2019semi}. However, along with the navigation process, users usually experience cybersickness due to mismatched visual and vestibular information in the brain: users visually perceive objects moving while the body is still in position~\cite{oman1990motion}. Therefore, this sensory conflict leads to sickness symptoms, such as headache, vomit, nausea, and sweating.

In this paper, we present a novel method to improve navigation experience in virtual environments. Our approach relies on online adaptation of navigation parameters based on pre-trained neural networks and laws from system control.

\subsection{Cybersickness evaluation}

Due to the individual susceptibility to cybersickness, the severity of the symptoms experienced distributes differently among users~\cite{Davis2014}. Many subjective evaluation methods were proposed in past research, such as the well-known \textit{Simulator Sickness Questionnaire} (SSQ)~\cite{Kennedy1993} that asks user to evaluate sixteen sickness symptoms after an immersive experience and classifies them into three categories: nausea, oculomotor and disorientation. However, researchers often stick to the overall cybersickness level. In this case, it might be meaningful to repeatedly ask participants a single question about their well-being, instead of asking about multiple symptoms. The \textit{Misery Scale} (MISC)~\cite{Bos2005,Bos2010} and the Fast Motion Sickness Scale (FMS)~\cite{Keshavarz2011} have been proposed accordingly for such convenience. However, subjective questionnaires are generally administered after users have performed an experiment; therefore they have to shift attention away from the experiment to personal feelings, leading to much disturbance on the resulting data~\cite{Dennison2016}.

Alternatively, biosignals are regarded as one of the most objective ways to represent individual differences (e.g., electrodermal activity (EDA), electroencephalography (EEG), heart rate variability (HRV), Eye tracking, \textit{etc.})~\cite{Dennison2016}. Recent work has demonstrated success in assessing and predicting cybersickness by involving these signals, especially thanks to deep learning approaches~\cite{padmanaban2018towards,Du2021,Hadadi2022}. For example, a pioneering work presented the encoding of EEG signals to the cognitive representation relative to cybersickness, and by transferring it to VR video-based deep neural networks, the authors can predict cybersickness without any EEG signal~\cite{Kim2019}. This work subtly integrates individual EEG information into the visual information, making it possible to predict individually different cybersickness. Islam et al.~\cite{islam2020automatic,Islam2021} put forward a multimodal deep fusion neural network that can take stereoscopic videos, eye-tracking, and head-tracking data as inputs for predicting the severity of cybersickness. Particularly, when using a combination of eye-tracking and head-tracking data, the authors found that their network can predict cybersickness with an accuracy of $87.77\%$, which has outperformed state-of-the-art research. Their approach gains strong feasibility for being used in current consumer-level HMDs having already integrated eye and head tracking sensors. However, the adoption of measurement devices such as EEG and eye tracking may be hindered by their intrusiveness and inconvenience for collection and analysis. Typically, the collection of EEG signals requires the experimenter to put several electrodes on the user's head~\cite{Krokos2022}. Despite EEG signals usually being noisy, such settings of real-time cybersickness evaluation may distract participants from the immersive experience. Recent progress in the development of wearable sensors provides insights for easy integration in immersive applications. One measurement that particularly attracted interest for several years in the VR community is the EDA (e.g., \cite{Plouzeau2018}). Indeed, the EDA signal can be easily recorded through a cheap electrical circuity, and is commonly regarded as a reflection of the sympathetic arousal~\cite{sharma2019audio}. In this work, we decided to opt accordingly for the EDA as a reliable cybersickness indicator.


EDA signals can be decomposed into two components: the skin conductance level (SCL) and the skin conductance response (SCR)~\cite{Aqajari}. SCL, associated with the tonic level of the EDA signal, changes slowly with a time scale of tens of seconds to minutes. Because of the differences in hydration, skin dryness, or autonomic regulation between respondents, SCL varies accordingly and can be significantly different among respondents. On the other hand, SCR, known as the phasic component of the EDA signal, rides on top of the tonic changes and demonstrates much faster variations. Alternations in SCR components of an EDA signal are observable as bursts or peaks in the signal. The phasic component is associated with specific emotionally arousing stimulus events (event-related SCRs, ER-SCRs). The rise of the phasic component can reach a peak within 1-5 seconds after the onset of stimuli~\cite{sharma2019audio}.

The SCL at the forehead and finger area can demonstrate correlation with cybersickness occurrences but not during the recovery stage~\cite{Golding1992}: the SCL will increase after users are exposed to visual stimulation~\cite{Koohestani2019}. According to the spectrometer measurement of water vapor produced by sweating, an increased SCL is linked to an increased sweating. After the termination of the visual stimuli, the conductive path through the skin remains open despite reduced or absent sweat gland activity. Additionally, past research has found that the SCR presents strong correlation with both the onset of and the recovery from cybersickness \cite{Golding1992}. SCRs collected from the forehead significantly indicate a sudden and sustained burst of activity preceding an increase in subjective cybersickness ratings. Though, SCRs gathered from the finger palmar site may not be related to the cybersickness, as the palm is less sensitive than the human forhead in both phasic and tonic levels~\cite{Koohestani2019}. Further past studies~\cite{Golding1997,Gavgani2017} confirmed that phasic changes of the skin conductance on the forehead can be used to measure the level of cybersickness. Last, it has been shown that the width of the SCR collected from the wristband could be an index of cybersickness~\cite{Magaki2019}, which further supports our approach to collect the EDA with a wristband sensor.

\subsection{Adaptive navigation}

Adaptive navigation in virtual reality refers to the use of techniques and algorithms that adjust the users' virtual experience based on their behavior and preferences. It is a way to enhance the interaction experience by customizing the virtual environment to their needs and abilities. The severity of cybersickness symptoms increases with time, and after some time spent in the VR, the sickness severity either begins to stabilize or decrease; therefore, adaptive navigation in the VR environment appears to be meaningful and deserve investigating~\cite{duzmanska2018can}.

Many strategies have been proposed in past research to mitigate cybersickness by adapting navigation settings. For example, fuzzy logic has been used to integrate three user factors (gaming experience, ethnic origin, age) to derive an individual susceptibility index to cybersickness \cite{wang2021fuzzy}. This work opens the possibility to adapt navigation settings based on individual characteristics. Fernandes and Feiner \cite{fernandes2016combating} explored the way to dynamically change the field of view (FOV) depending on the users' response to visually perceived motion in a virtual environment. As a result, users experience less cybersickness, without decreasing the sense of presence and minimizing the awareness to the intervention. However, Zielasko et al. failed to confirm the correlation between the reduced FOV and the severity of cybersickness, although a reduction of FOV allows the users to travel longer distances~\cite{zielasko2018dynamic}. To reduce the risk of cybersickness, Argelaguet and Andujar~\cite{Argelaguet2010} designed an automatic speed adaptation approach in which the navigation speed is computed on a predefined camera path using optical flow, image saliency and habituation measures. During navigation, users usually manipulate the speed based on the task and personal preferences, but they have to involuntarily adjust the speed frequently and unsmoothly, resulting in severe cybersickness. Hu et al.~\cite{hu2019reducing} carried out similar work for reducing cybersickness with perceptual camera control while maintaining original navigation designs. Considering the effect of speed on cybersickness, Wang et al.~\cite{wang2021speed} proposed an online speed protector to minimize the total jerk of the speed profile considering both predetermined speed and acceleration constraints, leading users to report less severity of cybersickness. Additionally, Freitag et al.~\cite{freitag2016automatic} developed an automatic speed adjustment method for travel in the virtual environment by measuring the informativeness of a viewpoint.

We believe that the evaluation and prediction of cybersickness is the first step, while the final objective is to reduce cybersickness through adaptive navigation. Therefore, through this work, we want to bridge the gap between evaluation and adaptation. A similar idea can be found in previous studies. Plouzeau et al.~\cite{Plouzeau2018} created an innovative method to adapt the navigation acceleration in real time based on the EDA signal, resulting in a significant decrease of cybersickness levels among users while maintaining the same task performance. Similarly, Islam et al. \cite{islam2021cybersense} designed a closed-loop framework to detect the cybersickness severity and adapt the FOV during navigation. The framework can collect the user's physiological data (e.g., HR, BR, HRV, EDA) with which cybersickness can be predicted, and based on the sickness severity, the system can apply dynamic Gaussian blurring or FOV reduction to the VR viewer.

These studies usually assume that the sickness severity has a linear correlation with the adapted settings, and therefore the navigation settings are adapted with a linear proportional function which is not confirmed to the best of our knowledge. However, naturally, the perception of visual stimuli could be a nonlinear process, and presetting a linear adaptation strategy may prevent from finding the optimal adaptive settings. Accordingly, we propose to use a Proportional-Integral-Derivative (PID) control to adapt the navigation settings, without requiring any assumption beforehand. In this work, we propose to use physiological signals, and particularly the electrodermal activity (EDA) that can be obtained from a wristband sensor in real time (here an Empatica E4 wristband\footnote{https://www.empatica.com/research/e4/}), to adapt navigation in real time, and therefore mitigate cybersickness. By involving the EDA, we expect to further incorporate individual differences into a customized navigation experience. We firstly use the phasic component of the EDA signal as a measurement of the cybersickness level in real time. Secondly, we formulate a mathematical relation based on the PID model with several unknown parameters to adapt the navigation acceleration, taking the sickness severity as input. Optimization is performed to determine the optimal parameters to output a nonlinear adaptation strategy.

Past work proposed a similar idea as ours in which the navigation acceleration is adapted from the real-time evolution of EDA \cite{Plouzeau2018}:

\begin{equation}\label{eq:jeremy}
    a(t_i)=a(t_{i-1}) - 0.5*\frac{dEDA(t_i)}{dt}
\end{equation}

where $a(t_i)$ and $a(t_{i-1})$ are the accelerations at two successive frames, $EDA(t_i)$ is the magnitude of the EDA signal at time $t_{i-1}$. This formulation is designed for an acceleration-based control scheme in which the joystick can control the longitudinal or rotational acceleration directly until reaching a speed limit. However, the user might fail to control the acceleration when $EDA(t_i)$ keeps increasing or decreasing, which is the reason why this model is not sufficient to mitigate cybersickness efficiently. We therefore propose another model, based on the PID control scheme, adding a second term in \autoref{eq:adaptive} to stabilize the acceleration. Our contributions are summarized as follows:

\begin{itemize}
  \item We develop a novel mathematical model for adapting the navigation acceleration based on the severity of cybersickness evaluated by the phasic component of the EDA signal.
  \item We propose and validate the use of deep neural networks (NN) in our studies. Deep NNs work as simulated users during the experiments, taking navigation as input and sickness as output. Such approach opens avenues to the development of intelligence in VR systems.
\end{itemize}

The paper is organized as follows. In \autoref{nav_model}, we will give a brief introduction to PID control and 1D convolutional neural networks, and we will present mathematically the proposed adaptive navigation model. In~\autoref{experiment}, we will demonstrate a feasible approach to find the optimal parameters for the adaptive model. Further, we will discuss the results and the limitations of the work in~\autoref{discussion}, before concluding.

\section{Adaptive Navigation Design}\label{nav_model}

\subsection{PID controller}

PID is the abbreviation of proportional-integral-derivative and is a control loop method with feedback widely used in industrial control systems and numerous applications needing constantly modulated control~\cite{wang2020pid}. A PID controller continuously computes an error value \textit{e(t)} representing the difference between the expected setpoint and the measured process value, and applies a correction determined by proportional, integral and derivative terms (denoted $P$, $I$, and $D$ respectively). The overall control function is given as,

\begin{equation}
    u(t)=K_P*e(t) +K_I*\int_{0}^\tau e(t)+K_D *\frac{de(t)}{dt}
\end{equation}

where, $u(t)$ is the output of the PID controller, $K_p$, $K_i$ and $K_d$ are non-negative coefficients for the proportional, integral and derivative terms respectively, $\tau$ is the time and $t$ is the integration variable. This controller contains three terms with different control purposes:

\begin{itemize}
    \item The proportional controller gives a feedback that is proportional to the error $e(t)$. If the error is large and positive, this term will also return a large and positive output, taking into account the coefficient $K_p$. However, it can not ensure that the system reaches the expected setpoint and maintains a steady-state error.
    \item The integral controller is involved to remedy the steady-state error. It integrates the historic cumulative value of the error until the error reduces to zero. However, the integral term decreases its output when a negative error appears, which will limit the response speed and influence the stability of the system.
    \item The derivative controller enables the system to estimate the future trend of the error based on its rate of change along time. It improves the stability of the system by compensating for a phase lag resulting from the integral term.
\end{itemize}

\subsection{1D convolutional neural network}


There exist multiple models of neural networks. Among them, 1D convolutional neural networks (CNN) can achieve competitive performance compared to for example long short term memory (LSTM) on certain sequence-processing problems, usually at a significantly cheaper computational cost. Recently, 1D CNNs have been applied for audio generation and machine translation, obtaining great success~\cite{Kiranyaz2021}. In this work, 1D CNNs will be used to process sequential data.

\subsection{Formulation of adaptive navigation}

The objective of this section is to deduce the mathematical formulations for the adaptive navigation technique based on the PID control system.

Let $f(t)$ denote the phasic component of the EDA signal at time $t$, and the objective is to stabilize $f(t)$, given as,

\begin{equation}\label{error_fun1}
E_f(t_i)= f(t_i)- f(t_{i-1})
\end{equation}

where $E_f(t_i)$ is the difference of the phasic component between the current time step $t_i$ and the previous time step $t_{i-1}$. In idle state, $f(t_i)$ is expected to be $0$ which means that there is no visual stimuli. Therefore, \autoref{error_fun1} can be simplified as,

\begin{equation}\label{error_fun2}
E_f(t_i)= - f(t_{i-1})
\end{equation}

\begin{figure}[!htp]
\centering
\includegraphics[width=0.99\columnwidth]{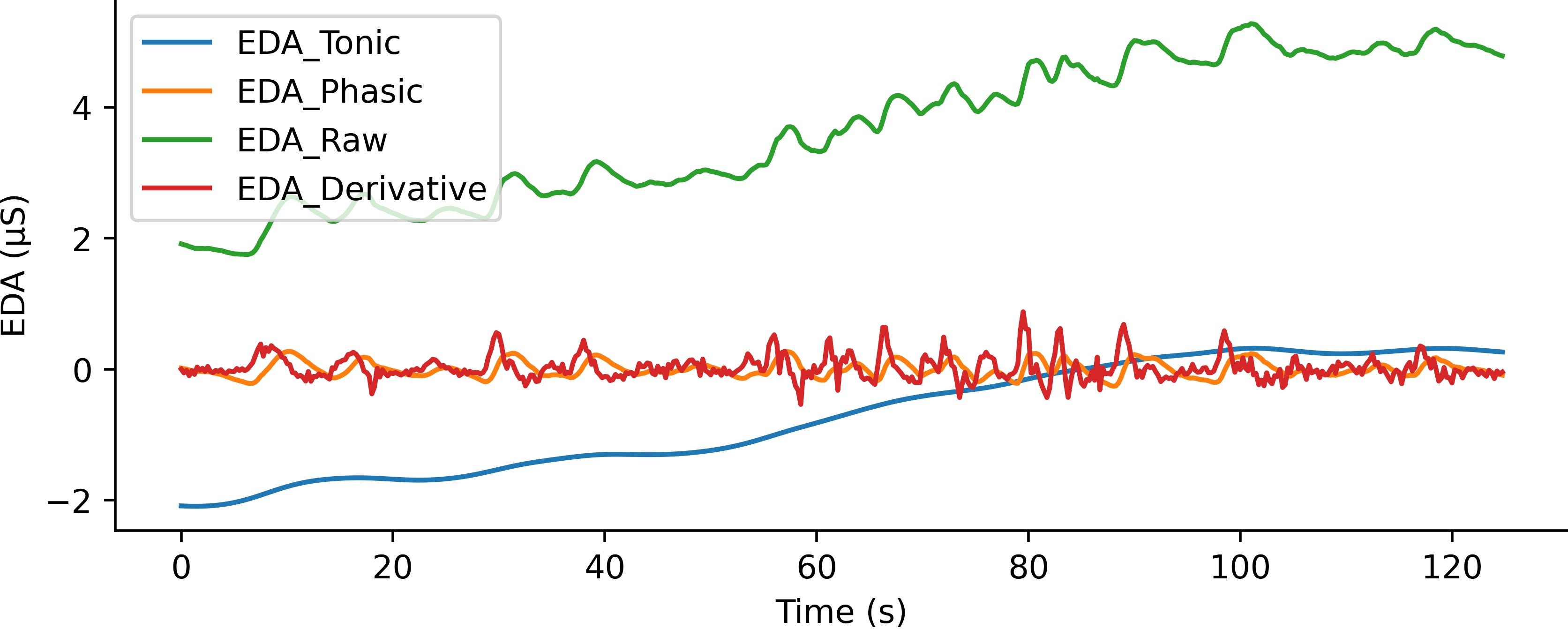}
\caption{Demonstration of one EDA signal including the phasic component, tonic component, and temporal derivative.}
\label{fig:eda} %
\end{figure}

As the EDA signal is decomposed into the phasic (SCR) and the tonic (SCL) component, it implies that the variation of the EDA signal is associated with the phasic and the tonic component. Knowing that the tonic component usually varies slowly and the phasic component varies rapidly overlying the tonic component~\cite{benedek2010continuous}, in practice, we can use the phasic component to approximate the variation of the EDA signal. Mathematically, the variation of the EDA signal by time is noted as a temporal derivative, i.e., $f=\frac{dEDA}{dt}$. As depicted in \autoref{fig:eda}, there exists a similar variation trend between the derivative of EDA and the phasic component of EDA.

As explained above, the SCR component is associated with arousing stimulus events. When a user is exposed to visual stimuli in immersive virtual environments, bursts or peaks appear. The severer the visual stimuli, the more salient the burst or the higher the peak. In other words, in an idle situation in which the user does not receive any visual stimuli, there should not be any observable bursts or peaks. Furthermore, it means that we should design a navigation technique in which the visual stimuli should not arouse excessive physiological responses. Our goal is then to optimize navigation to stabilize the SCR component of the EDA signal.


The navigation acceleration (both translational and rotational) $\mathbf{a}$ at time $t_i$ can be parameterized by the following algebraic expression,


\begin{equation} \label{eq:adaptive}
    \mathbf{a}(t_i)= \mathbf{a}(t_{i-1})+ \mathbf{\psi_a}\mathbf{E_a}(t_i) + \diag(\beta)  \psi_f \mathbf{E_f}(t_i)
\end{equation}
where

\begin{equation}\label{eq:psia}
\begin{aligned}
 \mathbf{\psi_a}(\cdot ) &=
  \begin{bmatrix}
    \psi_{a_l}(\cdot ) & 0   \\
     0& \psi_{a_r}(\cdot )
  \end{bmatrix}
 \\ & =
  \begin{bmatrix}
    K_{Pl}(\cdot) +K_{Il}\int_{0}^{\tau} (\cdot ) +K_{Dl} \frac{d(\cdot ) }{dt} &0  \\
    0 & K_{Pr}(\cdot)+K_{Ir}\int_{0}^{\tau} (\cdot ) +K_{Dr} \frac{d(\cdot )}{dt}
  \end{bmatrix}
\end{aligned}
\end{equation}

\begin{equation}\label{eq:psif}
\begin{aligned}
  \mathbf{\psi_f}(\cdot) &=
  \begin{bmatrix}
    \psi_f(\cdot) & 0   \\
     0& \psi_f(\cdot)
  \end{bmatrix}
 \\& =
  \begin{bmatrix}
    K_{Pf}(\cdot)+K_{If}\int_{0}^{\tau}(\cdot) +K_{Df}\frac{d(\cdot)}{dt} &0 \\
   0 & K_{Pf}(\cdot)+K_{If}\int_{0}^{\tau}(\cdot) +K_{Df} \frac{d(\cdot)}{dt}
  \end{bmatrix}
\end{aligned}
\end{equation}

\begin{equation}\label{eq:eaa}
  \mathbf{E_a}(t_i) =
  \begin{bmatrix}
    E_{a_l}(t_i)   \\
     E_{a_r}(t_i)
  \end{bmatrix}
  =
  \begin{bmatrix}
    a_{le}(t_i)-a_l(t_i)  \\
    a_{re}(t_i)-a_r(t_i)
  \end{bmatrix}
\end{equation}

\begin{equation}\label{eq:eaf}
  \mathbf{E_f}(t_i) =
  \begin{bmatrix}
    E_f(t_i)   \\
     E_f(t_i)
  \end{bmatrix}
  =
  \begin{bmatrix}
    - f(t_{i-1})  \\
     - f(t_{i-1})
  \end{bmatrix}
\end{equation}

\begin{equation} \label{eq:diag}
  \diag(\beta) =
  \begin{bmatrix}
    \beta_l & 0  \\
     0 & \beta_r
  \end{bmatrix}
\end{equation}

In \autoref{eq:adaptive}, the second term  $\mathbf{\psi_a}\mathbf{E_a}(t_i)$  ensures that the acceleration can vary around an expected value. Indeed, if the third term $ \psi_f \mathbf{E_f}(t_i)$ keeps varying monotonously, the acceleration would also vary monotonously and reach an extremum. The third term implies then the adaptive quantity due to the visual stimulus or physiological response. $\diag(\beta)$ represents diagonal coefficient matrices used to balance the importance between longitudinal and rotational motion in \autoref{eq:adaptive}.

$a_{le}$ is the expected longitudinal acceleration; $a_{re}$ is the expected rotational acceleration; $a_l$ is the measured longitudinal acceleration; $a_r$ is the measured rotational acceleration. Both expected accelerations are $0$.

$\mathbf{\psi_a}(\cdot )$ and $\mathbf{\psi_f}(\cdot)$ are the PID operators. $\psi_{a_l}(\cdot)$ and $\psi_{a_r}(\cdot)$ are elements of $\mathbf{\psi_a}(\cdot )$ with the following coefficients: $K_{Pl}$, $K_{Il}$, $K_{Dl}$, and $K_{Pr}$, $K_{Ir}$, $K_{Dr}$, acting on the longitudinal and rotational accelerations respectively; they are used to ensure the accelerations to be around the expected values. $\psi_{f}(\cdot)$ is the element of $\mathbf{\psi_f}(\cdot)$ with the following coefficients: $K_{Pf}$, $K_{If}$, $K_{Df}$, acting on the phasic component of the EDA signal. $\mathbf{E_a}(t_i)$ and $\mathbf{E_f}(t_i)$ are the errors between the expected steady states and the measured states at time step $t_i$. The errors will be substituted to the PID operators to compute the corrections.

To summarize, an adaptive navigation system can be described by \autoref{eq:adaptive} and the supplementary equations: \autoref{eq:psia}, \autoref{eq:psif}, \autoref{eq:eaa}, \autoref{eq:eaf}, and \autoref{eq:diag}. The inputs of the system at time step $t_i$ are the measured longitudinal and rotational accelerations, and the phasic component of EDA, and the outputs of the system are the corrected accelerations in both directions.

This parameterized model features the following eleven coefficients: $K_{Pl}$, $K_{Il}$, $K_{Dl}$, $K_{Pr}$, $K_{Ir}$, $K_{Dr}$, $K_{Pf}$, $K_{If}$, $K_{Df}$, $\beta_l$ and $\beta_r$. As the objective of adaptation is to mitigate cybersickness, the nontrivial question is: can we find the optimal parameters to adapt the acceleration allowing to reduce cybersickness?

\begin{figure*}[!htp]
\centering
\includegraphics[width=0.75\textwidth]{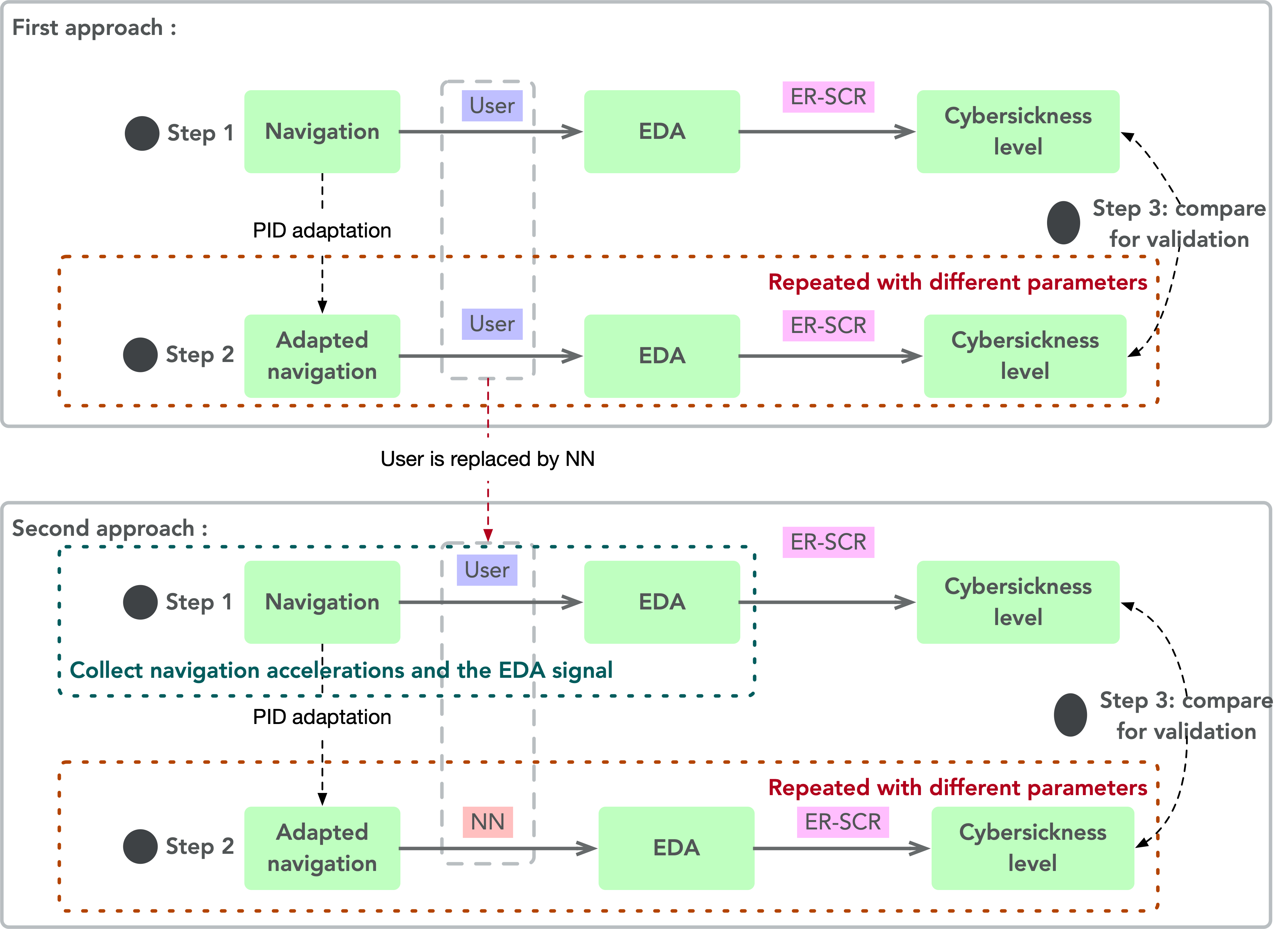}
\caption{Strategy to find the optimal parameters that can mitigate cybersickness in the adaptive model. }
\label{fig:view} %
\end{figure*}

We can see two possibilities to answer this question, depicted in \autoref{fig:view}. The first approach is to assign different values to these coefficients and perform user studies to determine the best one that can mitigate cybersickness. As there are eleven coefficients and each of them is independent to the others, the combination of optimal coefficients reaches at least hundreds or thousands of groups. To prove that a specific group of coefficients can reduce cybersickness with a statistical significance, dozens of participants are required to evaluate the adaptation system, with much time needed to complete all the experiments. In general, three steps are required for this approach that we will call classical:
\begin{enumerate}
    \item A group of users navigates through a virtual environment without adaptive navigation and evaluates the corresponding cybersickness level according to the EDA signal afterwards.
    \item The same group of users navigates through the same virtual environment with the PID adapted navigation. In this step, we have to manually find appropriate values for the different parameters.
    \item After evaluating the cybersickness level, a comparison between both approaches is perform to determine potential significantly reduced cybersickness levels.
\end{enumerate}

The recent strong development of artificial intelligence in various domains has raised interest in developing intelligent systems that can predict phenomena at a price of less efforts for developers and end-users. Although considering human participants cannot be dismissed, the release of more and more efficient neural network (NN) algorithms represents a formidable opportunity to introduce them into VR applications and pave foundations for the development of more efficient and individualized VR. Particularly, as in the classical approach above, parameter tuning can be highly time-consuming and tiring for participants, using NNs to simulate users may be an interesting alternative to explore. In fact, in the issue considered in this work, 
the function of users is to map the motion profiles during navigation to the corresponding EDA signal through which we can compute the cybersickness level. Fortunately, neural network (NN) models can work for this purpose as they are widely regarded as nonlinear fitters. Therefore, in this paper, we propose to investigate whether, and if so, how, with NNs, we can determine the performance of the adapted navigation with less user studies; and here comes the second approach: we propose to use NNs to simplify user studies and evaluate the performance of the different coefficients. This approach includes three steps:

\begin{enumerate}
    \item The first step is similar to that of the classical approach. We collect from past user studies longitudinal and rotational accelerations and EDA signals during navigation in an immersive environment. The collected data are then used to train the NN model.
    \item With the trained NN model, any navigation acceleration can be ingested to output the corresponding EDA signal. Therefore, in this step we can replace users by the trained NN model, allowing not to conduct a vast number of user studies. The focus of the work becomes then to search for appropriate parameters that can mitigate cybersickness, which corresponds exactly to a parameter optimization problem. In mathematics, parametric optimization can be solved with different methods that have already been implemented in open-source softwares such as \textit{Optuna}~\cite{optuna_2019}. Once the objective of the optimization is determined, \textit{Optuna} can find the optimal parameters of our model through a Bayesian optimization, especially sequential model-based optimization with Tree-Structured Parzen Estimator.
    \item Similar to that in the classical approach, the third step investigates whether adapted navigation can reduce cybersickness by comparing the artificially generated ER-SCR feature of the EDA signal to that coming from non-adapted navigation.
\end{enumerate}

Compared to the classical approach, the benefits of the second approach are: (1) replacing users by an NN, which alleviates the challenge of recruiting numerous participants; (2) transposing the adaptation problem to an optimization problem that can be solved with existing open-source software. In this case, the objective of the experiment is to collect enough data to train the NN model.

\section{Data Collection and Parameters Computation}\label{experiment}

\subsection{Data acquisition}

To avoid performing excessive user tests, we had to train an NN model that can map the navigation behavior (acceleration in this context) to the corresponding EDA signal. However, cutting down the number of user tests does not imply that we can completely get rid of them, we still need to collect enough data to train a high-quality NN model. Hence, we carried out a user experiment to collect the required data.

\subsubsection{Participants}

We invited $53$ participants ($M_{age}=26.3$, $SD_{age}=3.3$, Females: $26$) from the local city to participate in a navigation task in an immersive environment. To obtain much more samples, all participants were asked to participate three times on three different days, hence we collected $159$ samples. They were rewarded with different gifts afterwards. Upon arrival, they were asked to fill one pre-exposure questionnaire to investigate on their health conditions and experience in playing games and using VR devices. From this questionnaire, no participants reported any health issues that would affect the experiment results. A consent form was signed by participants.

\subsubsection{Task design}

\begin{figure*}[tb]
\centering
\includegraphics[width=0.9\textwidth]{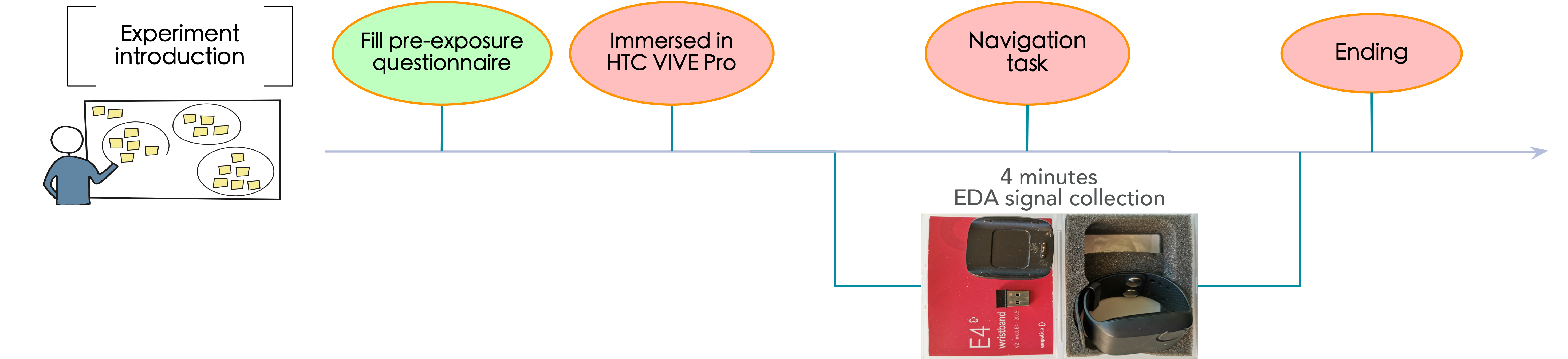}
\caption{Flowchart for the experiment including user navigation with HTC Vive Pro and data collection with Empatica E4 wristband.}
\label{fig:experiment} %
\end{figure*}

\begin{figure}[tb]
\centering
\includegraphics[width=0.7\columnwidth]{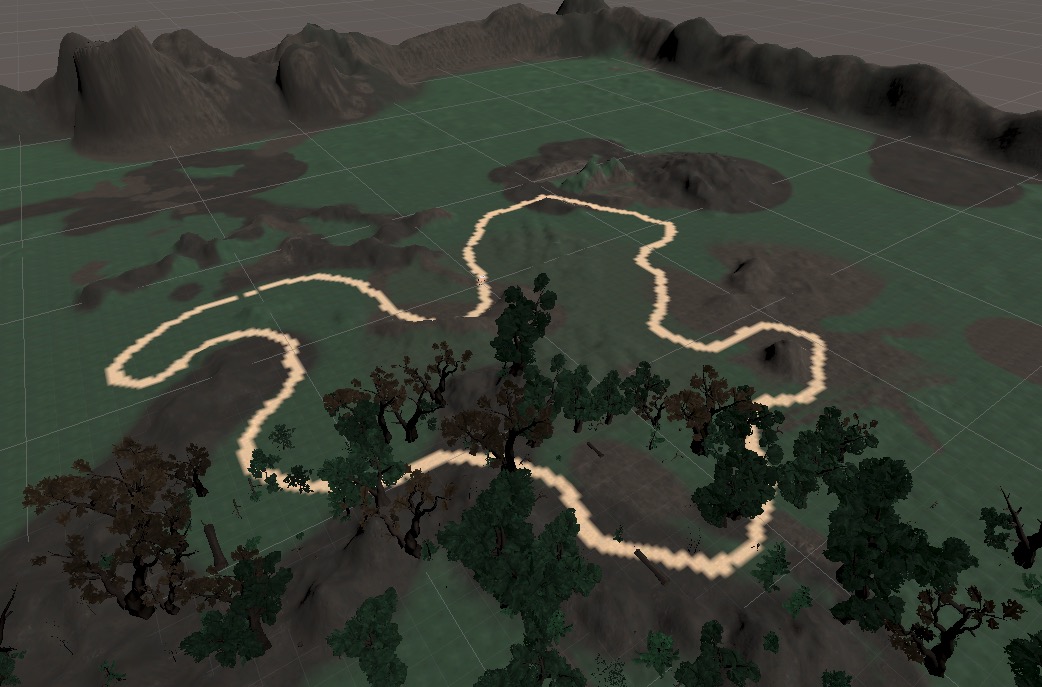}
\caption{Virtual scenario in which the participants navigate along the highlighted path.}
\label{fig:scenario} %
\end{figure}

The general experimental procedure is presented in \autoref{fig:experiment}. The whole experiment was carried out using an HTC Vive Pro head-mounted display.

\begin{enumerate}
    \item Before the test, we gave the participants a brief introduction about how to control navigation with the HTC Vive Pro hand controllers. Due to the occurrence of cybersickness, they were allowed to terminate the experiment whenever they felt sick or severe discomfort.
    \item The experimenter put an HTC Vive Pro on the participants' head and an Empatica E4 wristband on one participants' arm. The Empatica E4 can sample EDA at a frequency of $\SI{4}{\hertz}$ and the EDA signal is sent during navigation to a processing computer through Bluetooth.
    \item The participants were immersed in the virtual environment displayed in \autoref{fig:scenario} and started to navigate following the trajectory highlighted in brown. The user could control the motion in different directions through the touchpad on the HTC Vive Pro hand controller. Together with the EDA signal, the longitudinal and rotational navigation accelerations were recorded synchronously. 
    \item The navigation task continued for four minutes, and the participants were removed from the head-mounted display at the end.
\end{enumerate}

\subsection{Model architecture}

\begin{figure*}[!htp]
\centering
\includegraphics[width=0.9\textwidth]{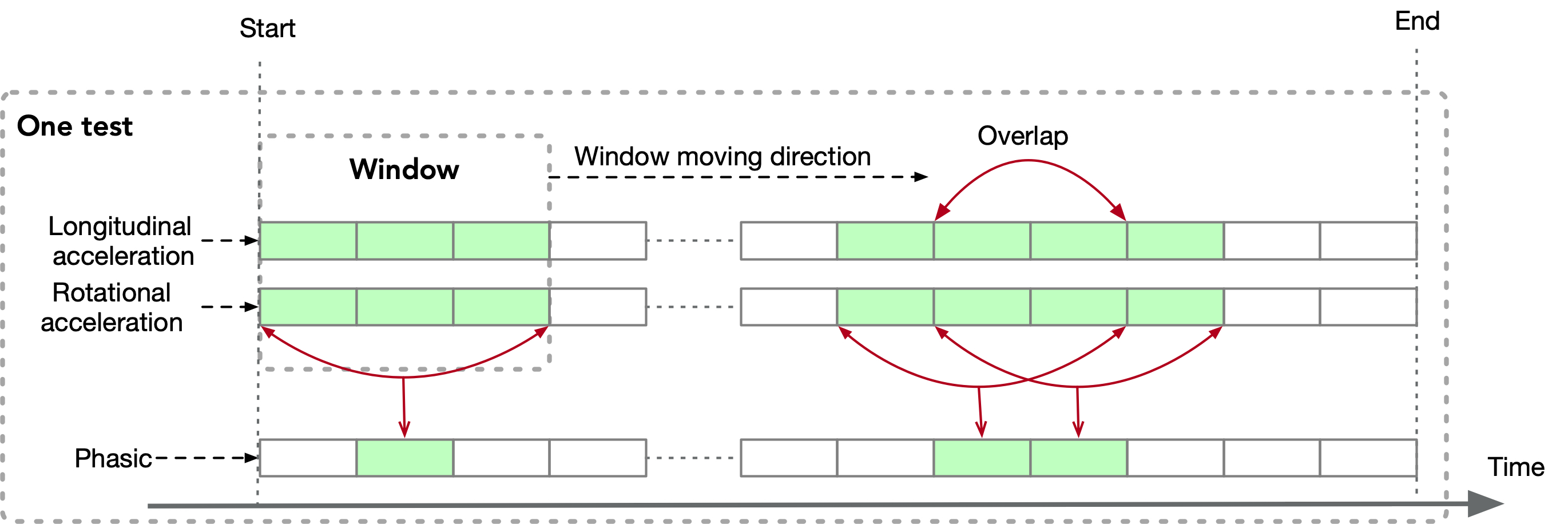}
\caption{Schematic representation of the data collected from one user session including the longitudinal and rotational accelerations, and the phasic component of EDA. Note that both accelerations were computed from the navigation speeds, and the phasic component of EDA was also preprocessed by the \textit{Neurokit2} package. }
\label{fig:dataset} %
\end{figure*}


During the experiment, we collected three signals with the same starting and ending times: the longitudinal acceleration, the rotational acceleration, and the EDA signal. As the phasic component is associated with arousing stimulus events, the NN model should link the navigating accelerations to the phasic component of EDA. Therefore, we extracted the phasic component from the original EDA thanks to \textit{Neurokit2}~\cite{Makowski2021neurokit}, a Python toolbox for neurophysiological signal processing. \autoref{fig:dataset} represents a schematic representation of the data recorded from one participant session. Individual differences leading to different magnitudes in the phasic response could make the model difficult to train; thus we normalized all the data to range between zero and one.

\begin{figure*}[ht]
\centering
\includegraphics[width=0.4\columnwidth,angle=90]{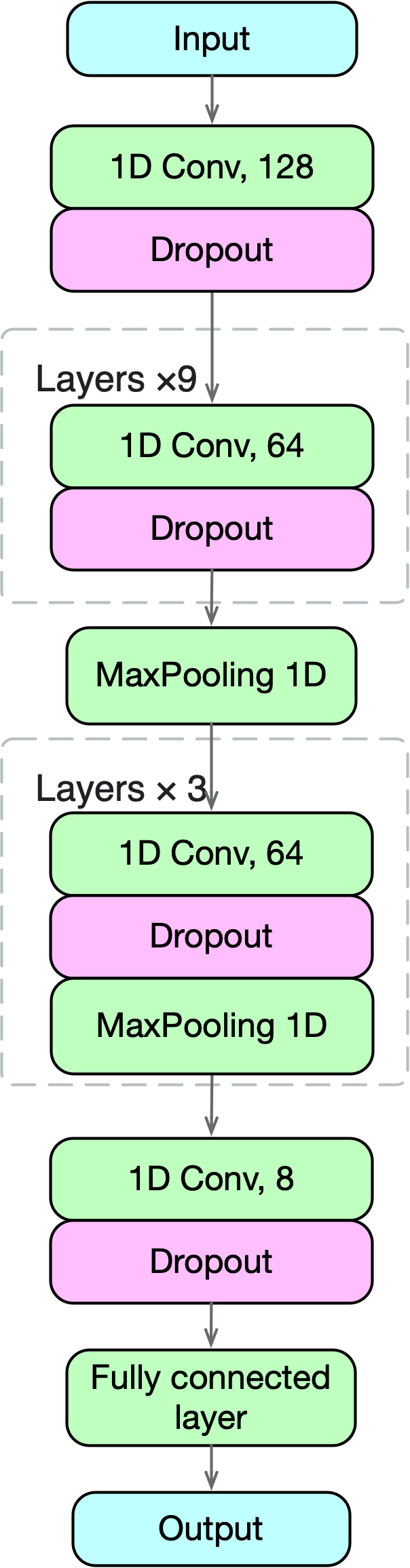}
\caption{Architecture of the deep neural network. The network is composed of fourteen 1D convolutional layers and one fully connected layer; the kernel size for the convolutional is 3.}
\label{fig:model} %
\end{figure*}

To allow the model to learn the local relationship between the acceleration and the phasic signal more easily, we used a moving window on the signal and reformulated the data structure. The current time clip of the phasic signal was associated with the accelerations from the previous time clip and the next time clip. For example, with a clip length of $\SI{1}{\second}$, to predict the phasic signal between $\SI{1}{\second}$ and $\SI{2}{\second}$, we used the acceleration data from $\SI{0}{\second}$ to $\SI{3}{\second}$; to predict the phasic signal between $\SI{2}{\second}$ and $\SI{3}{\second}$, we used the acceleration data from $\SI{1}{\second}$ to $\SI{4}{\second}$. The duration of the window clip was considered as a hyper-parameter for the NN model. In total, we collected $159$ pairs of data\footnote{One pair of data includes longitudinal and rotational accelerations and the corresponding EDA signal from one user session; one pair can be regarded as one data sample} among which we used $119$ pairs to train the model, and the rest of them ($40$) to test whether the model could reduce the level of cybersickness. With $119$ pairs of data, we obtained $90530$ clips as the training set, and $22633$ as the testing set for the NN model.

The proposed model was implemented in the Tensorflow framework on an Nvidia GeForce RTX 2070 graphic card. Adam was chosen as the optimizer. Other hyper-parameters in the NN model (dropout rate, learning rate, the number of convolutional layers, the number of epochs, batch size) were optimally determined by \textit{Optuna}. It took approximately three hours to train the model, and \textit{Optuna} spent around ten days to find the optimal hyper-parameters inside the searching space. Eventually, \textit{Optuna} reported the best prediction accuracy in terms of the mean absolute error (MAE) is $0.015$ (1D CNN) and $0.058$ (LSTM). Therefore, we used 1D CNN in this work considering the lower loss error.

The hyper-parameters of the 1D CNN model are reported in \autoref{tab:nn}. The model structure is given in \autoref{fig:model}.

\begin{table}[htp]
\centering
\caption{Settings of the hyper-parameters obtained from \textit{Optuna}. }\label{tab:nn}
\begin{tabular}{cc}
\toprule
Phasic signal length          & $\SI{2.25}{\second}$   \\ \hline
Dropout rate        & 0.00099                          \\ \hline
Learning rate       & 0.000288                         \\ \hline
\# Convolutional layers & 14                               \\ \hline
Epoch               & 1800                             \\ \hline
Batch size          & 256                              \\
\bottomrule
\end{tabular}
\end{table}

\subsection{Computing the adaptive coefficients}

With the previously determined hyper-parameters, we obtained the best NN model that can map the accelerations to the phasic component of the EDA signal. At this stage, we can further find the optimal adaptive coefficients using the $40$ pairs of data mentioned above. The idea is to employ \textit{Optuna} again to use the number of ER-SCR as the optimization objective, and search for the best adaptive coefficients that can reduce the number of ER-SCR. For the non-adapted navigation, the phasic component of EDA is obtained from the data collected during the user study, while for the adapted navigation, the phasic component of EDA is predicted from the pre-trained NN model. The detailed steps of the process to find the optimal coefficients of the adapted navigation model (the eleven coefficients introduced in \autoref{nav_model}) are,

\begin{figure*}[!htp]
\centering
\includegraphics[width=0.7\textwidth]{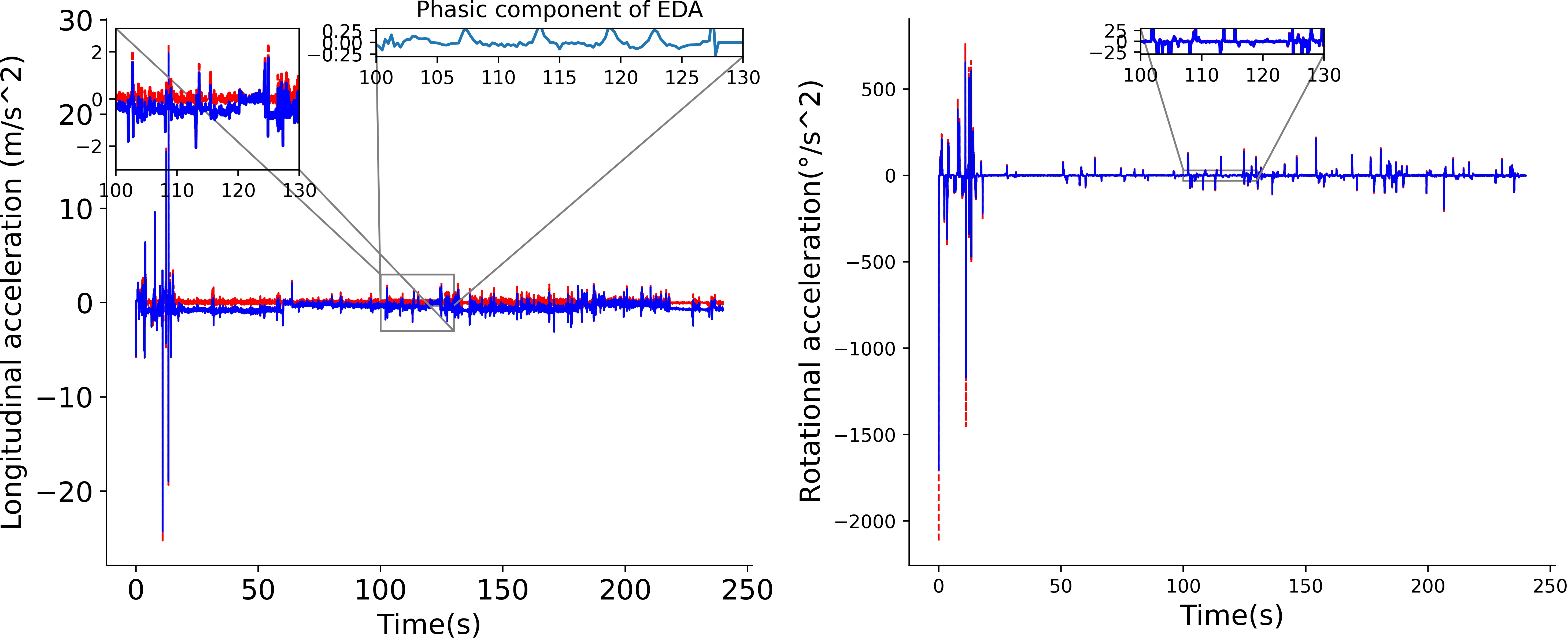}
\caption{Comparison between non-adapted (in red) and adapted (in blue) longitudinal (left) and rotational (right) accelerations. The adapted accelerations are determined by the phasic component of EDA (zoomed part in the left graph).}
\label{fig:adaptedSpeed} %
\end{figure*}



\begin{enumerate}
    \item With the $40$ pairs of data, we can compute the number of ER-SCR for the non-adapted navigation of each sample, denoted by a vector $\mathbf{N}_{raw}$, used as a baseline to be compared with the adapted one.

    \item \textit{Optuna} will randomly choose values for the eleven unknown coefficients. With the model proposed in~\autoref{eq:adaptive}, we can compute a value for the adapted navigation accelerations based on the non-adapted one.

    \item After obtaining the adapted accelerations, we can use the pre-trained NN model to compute the phasic component of EDA. As the pre-trained model only reads and returns the clipped data (as shown in~\autoref{fig:dataset}), an additional step will reconstruct the phasic component obtained from multiple clipped data to a single sequence, and further compute the number of ER-SCR which is denoted by another vector $\mathbf{N}_{adapted}$. 
    
    \item We compute the difference ($\mathbf{N}_{raw} - \mathbf{N}_{adapted}$) to record the reduction of the number of ER-SCR, and the result is denoted by one vector $\mathbf{N}$ containing both positive and negative numbers. A positive number means that the person will experience reduced cybersickness, and a negative number denotes increased cybersickness. The goal during this optimization process is to have more positive numbers than negative ones, i.e., maximizing the percentage of positive numbers among the 40 samples. As the \textit{Neurokit2} package provides diverse methods for computing the ER-SCR, we employed three methods from Kim~\cite{kim2004emotion}, Gamboa~\cite{gamboa2008multi}, and \textit{Neurokit2}~\cite{Makowski2021neurokit} to compute the percentage of positive numbers. Next, another group of values will be randomly chosen for the coefficients and investigation will be performed to further increase the percentage of positive numbers, denoted by $P_{pn}$. The following formulation summarizes the optimization process operated by \textit{Optuna} to find the optimal PID coefficients,
    
    \begin{equation}
    \begin{aligned}
        \max_{\text{PID coefficients}} P_{pn} = & \underbrace{\text{Percentage of positives}}_{\text{from Kim~\cite{kim2004emotion}}}
        \\& + \underbrace{\text{Percentage of positives}}_{\text{from Gamboa~\cite{gamboa2008multi}}}  
        \\& + \underbrace{\text{Percentage of positives}}_{\text{from \textit{Neurokit2}~\cite{Makowski2021neurokit}}}
    \end{aligned}
    \end{equation}
    
    \item Steps 2, 3 and 4 are repeated until the $P_{pn}$ reaches a stable value. Once convergence is achieved, the computed coefficients are considered as the optimal ones, in our case, the ones reported in \autoref{tab:pid}.
\end{enumerate}

\begin{table}[!htp]
\centering
\caption{Optimal coefficients of the adaptive model obtained from \textit{Optuna}. }\label{tab:pid}
\begin{tabular}{cc}
\toprule
$K_{Pl}$    &   0.0113    \\ \hline
$K_{Il}$    &   0.0065    \\ \hline
$K_{Dl}$    &   0.0137   \\ \hline
$K_{Pr}$    &   0.0098    \\ \hline
$K_{Ir}$    &   0.0012    \\ \hline
$K_{Dr}$    &   0.0011    \\ \hline
$K_{Pf}$    &   0.0730    \\ \hline
$K_{If}$    &   0.2283    \\ \hline
$K_{Df}$    &   0.3724     \\ \hline
$\beta_l$    &  0.0017     \\ \hline
$\beta_r$    &  0.0012     \\
\bottomrule
\end{tabular}
\end{table}

\subsection{Results}

Based on the methodology presented above, a theoretical validation test was conducted in which the trained NN replaced users and adapted navigation was simulated numerically.

\autoref{fig:adaptedSpeed} demonstrates the difference between non-adapted and adapted accelerations in the longitudinal and rotational directions for one test in the scenario presented in \autoref{fig:scenario}. During non-adapted navigation, the acceleration might reach extremely large magnitudes because the user moves rapidly. Adaptive navigation could compensate for such variation.

\begin{table}[h]
\centering
\caption{The majority of samples (total is 40) report significantly decreased numbers of ER-SCR and $MSDV$ with our adaptive model. }\label{tab:results}
\scalebox{0.9}{
\begin{tabular}{cccccc}
\toprule
\textbf{Methods} & \textbf{\begin{tabular}[c]{@{}c@{}}Positive\\ number\end{tabular}} & \textbf{Percentage} & \textbf{$\chi^2$} & \textbf{P-value} & \textbf{$\phi$} \\ \hline
Kim2004          & 36                                                                 & 90\%                & 25.6              & \textless{}.01   & 0.8             \\
Gamboa2008       & 29                                                                 & 72.5\%              & 8.1               & \textless{}.01   & 0.45            \\
Neurokit2         & 36                                                                 & 90\%                & 19.6              & \textless{}.01   & 0.7             \\ \hline
$MSDV_l$         & 40                                                                 & 100\%               & 40.0              & \textless{}.01   & 1.0             \\
$MSDV_r$         & 40                                                                 & 100\%               & 40.0              & \textless{}.01   & 1.0            \\
\bottomrule
\end{tabular}
}
\end{table}

To validate whether the adapted model with the coefficients computed in \autoref{tab:pid} can mitigate cybersickness, we compared the non-adapted navigation and the adapted one from a statistical viewpoint. We used the \textit{Neurokit2} package to compute the number of ER-SCR in each navigation modality. The significance level was set to $.05$.


Results in \autoref{tab:results} reveal that the adapted model can reduce the number of ER-SCR. According to the evaluation from the method of Kim~\cite{kim2004emotion} and \textit{Neurokit2}~\cite{Makowski2021neurokit}, $90\%$ of the total samples presented a significant reduced number of ER-SCR, ($\chi^2(1,N=40)=25.6,p<.01,\phi=.80$) and ($\chi^2(1,N=40)=8.1,p<.01,\phi=.45$) respectively, whereas only $72.5\%$ from the method of Gamboa~\cite{gamboa2008multi} provided significant results ($\chi^2(1,N=40)=19.6,p<.01,\phi=.70$).

Additionally, to further validate the performance of the proposed adaptive navigation model, still using the samples above, we computed the motion sickness dose value ($MSDV$) which is regarded as an objective cybersickness indicator~\cite{so1999search,kilteni2012sense,aykent2014motion}: a small $MSDV$ indicates less severity of cybersickness. The $MSDV$ can be computed with the following formula:

\begin{equation}
    MSDV=\sqrt[n]{\int_0^T a^{n}(t)\rm{d}t}
\end{equation}

where $a$ is the navigation acceleration ($\rm{m/s^2}$), $T$ is the whole navigation time, and $n$ equals to $2$ here. The PID-adapted acceleration could significantly reduce the $MSDV$ along both longitudinal or rotational directions for all samples ($\chi^2(1,N=40)=40,p<.01,\phi=1.0$).



\section{Discussion}\label{discussion}

Our model incorporates all accelerations (translational and rotational) and coefficients into an algebraic form. In \autoref{eq:jeremy}, the coefficient $0.5$ was derived empirically considering the physiological reaction time of the EDA \cite{Plouzeau2018}, while in our model, all coefficients were optimized based on existing datasets constructed from past user studies.

Our adaptive navigation model was used to adapt the navigation acceleration, while it is worth noting that it could be used to adapt other navigation settings, e.g., field-of-view (FOV) vignetting~\cite{fernandes2016combating} or geometry deformation~\cite{lou2019reducing}. To adapt different navigation parameters, we can replace the acceleration in \autoref{eq:adaptive} by the corresponding parameters and then follow the same procedure as shown in this study to find the best parameters. The objective of this work was not to compare the performance of different adaptive navigation settings for mitigating cybersickess, but to design a model of an adaptive strategy, and showcase it to one navigation parameter. For example, Islam et al. \cite{islam2021cybersense} developed a closed-loop framework to detect cybersickness and adapt the FOV accordingly; their work has gained advantage by using a deep LSTM neural network to predict real-time cybersickness with physiological data including heart rate, heart rate variability, EDA and breathing rate, while our work only used the phasic component of EDA to detect cybersickness. Therefore, we can legitimately wonder whether involving other physiological signals can improve the accuracy of adaptation. Even so, the main difference here is that after getting a feedback signal (i.e., the level of cybersickness), our model can make use of the proportional, integral and derivative components for adaptation, that can process problems with high nonlinearities.

We used a framework called \textit{Optuna} twice in this work with different intentions. \textit{Optuna} is an automatic hyperparameter optimization framework designed for optimizing an NN model. First, to avoid performing massive user studies, we needed an NN model to link navigation accelerations to the phasic component of EDA, and \textit{Optuna} could help find the optimal parameters for the NN model. Although we came up with an NN model to reproduce human cognition, it could be a promising method in many user interaction design: since the NN model has been trained on the navigation data from real user experiments, our model can predict the user response (e.g., EDA) according to the visual stimuli, allowing a designer to improve the interaction interface based on the predicted response~\cite{yang2020measuring}. Second, after proposing the adaptive model, we had to determine the optimal parameters allowing to significantly mitigate the cybersickness level; therefore, we employed \textit{Optuna} again for this parameter optimization problem.


Physiological measures have been praised as objective indicators for cybersickness and affective experience~\cite{Wang2019,kaneko2021comparing}. There was no need to employ subjective cybersickness measurements, such as SSQ, FMS, and MISC because the computational model evaluated cybersickness from an implicit point in real time instead of explicit subjective or verbal feedback which can only evaluate cybersickness for a certain duration. However, physiological signals have opened the possibility to continuously measure cybersickness, although the sensitivity of these signals to visual stimuli might limit the performance. For example, the duration of SCR windows usually varies from 1 to 5s after the onset of stimulus, and there might be overlapping in two subsequent ER-SCRs if the recovery time is large than the inter-stimulus interval, leading to distortion of the ER-SCR~\cite{sharma2019audio}.

Despite promising results achieved in simulation, more validation studies are required. The optimal parameters for adaptive navigation were found from a theoretical point, but we believe that an additional user study to compare the non-adapted and adapted navigation can further confirm the effectiveness of our approach. Put differently, the lack of user studies might weaken the reliability of the parameters computed. An experimental validation after the theoretical study not only can validate the performance of the adaptive parameters, but also can help define the search range in \textit{Optuna} and strengthen the relationship between the model parameters, thus it can bring benefits to the tuning process. We intend to carry out such studies in future research, and at this stage we encourage readers to focus on the adaptive model apart from finding the optimal parameters. In addition, our model involved eleven parameters in order to adapt navigation. Future research can investigate deeply on the relationship between these parameters and simplify the model to have less parameters.

\section{Conclusion}\label{conclusion}

We proposed a pioneering mathematical model for adaptive navigation in virtual environments by integrating a PID controller, with a long-term vision that immersive experience should be individualized. The premise to run this model successfully requires the system to detect cybersickness accurately; otherwise the adaptive power from the PID controller is weakened. Many adaptive VR systems have been focusing on the detection and evaluation of cybersickness in immersive environments, while we paid more attention to utilizing the cybersickness to optimize the navigation settings backward. The work is a theoretical paper with a solid simulated validation based on the number of ER-SCR and $MSDV$. Our contribution is to lay down the foundations of intelligent VR, in which a VR system can act as an assistant to help users perform better, which justifies the need to build computational models of cybersickness involving the generation of artificial data through AI. The pandemics has further been a great facilitator to introduce such approach. Although we found optimal adaptive coefficients thanks to simulation in \textit{Optuna}, we are planning to run more user studies to further improve its performance.





\acknowledgments{%
The authors show their deepest gratitude to the experiment participants. 
}

\bibliographystyle{abbrv-doi-hyperref}

\bibliography{Ismar_PID/reference}








\end{document}